\documentclass[a4paper,12pt]{article}

\usepackage{amsmath}
\usepackage{bm}
\usepackage{amssymb}
\usepackage{epsfig}
\usepackage{url}
\usepackage{color}
\usepackage[utf8]{inputenc} 
\usepackage{amsmath}
\usepackage{amssymb}
\usepackage{slashed}

\usepackage{cancel}
\usepackage{textcomp}
\usepackage{calc}

\usepackage{color}
\usepackage{soul,xcolor}
\allowdisplaybreaks

\catcode`\@=11
\def\lsim{\mathrel{\mathpalette\@versim<}}
\def\gsim{\mathrel{\mathpalette\@versim>}}
\def\@versim#1#2{\vcenter{\offinterlineskip
\ialign{$\m@th#1\hfil##\hfil$\crcr#2\crcr\sim\crcr } }}
\catcode`\@=12

\makeatletter
\def\lsim{\mathrel{\mathpalette\@versim<}}
\def\gsim{\mathrel{\mathpalette\@versim>}}
\def\@versim#1#2{\vcenter{\offinterlineskip
        \ialign{$\m@th#1\hfil##\hfil$\crcr#2\crcr\sim\crcr } }}

\makeatother
\newcommand{\vs}[1]{\vspace{#1 mm}}

\def\rmF{{\rm F}}

\def\rmT{{\rm T}}

\def\bfx{{\boldsymbol x}}
\def\bfy{{\boldsymbol y}}
\def\bfp{{\boldsymbol p}}

\def\bfq{{\boldsymbol q}}

\def\Re{{\rm Re}\,}
\def\Im{{\rm Im}\,}
\def\calL{{\cal L}}

\def\nn{\nonumber\\}

\def\half{\frac12}
\def\sqr#1#2{{\vcenter{\hrule height.#2pt
      \hbox{\vrule width.#2pt height#1pt \kern#1pt
          \vrule width.#2pt}
      \hrule height.#2pt}}}
\def\bra#1{\left\langle{#1}\right|}
\def\ket#1{\left|{#1}\right\rangle}
\def\VEV#1{\left\langle{#1}\right\rangle}

\def\eps{\epsilon_g\,}
\def\myLlimit#1{\ \ 
        \mathrel{\mathop{\kern0pt \Llongrightarrow }\limits_{#1}}\ \ }
\def\Llongrightarrow{\relbar\kern-3pt\relbar\kern-3pt\relbar\kern-3pt\relbar\kern-3pt\longrightarrow}

\makeatletter

\begin{document}

\hspace{11cm}
YITP-24-20


\medskip
\renewcommand{\thefootnote}{\fnsymbol {footnote}}

\begin{center}
{\Large\bf  
Anti-Instability of Complex Ghost}
\vs{10}

{\large
Jisuke Kubo,$^{1,2,}$\footnote{e-mail address: kubo@mpi-hd.mpg.de}
and Taichiro Kugo,$^{3,}$\footnote{e-mail address: kugo@yukawa.kyoto-u.ac.jp}
} \\
\vs{5}

$^1$
{\it 
Max-Planck-Institut f\"ur Kernphysik(MPIK), Saupfercheckweg 1, 69117 Heidelberg, Germany}

$^2$
{\it 
Department of Physics, University of Toyama, 3190 Gofuku, Toyama 930-8555, Japan
}

\vs{3}
$^3$
{\it Center for Gravitational Physics
and Quantum Information,
\\ Yukawa Institute for Theoretical Physics,
Kyoto University, Kyoto 606-8502, Japan}

\vs{10}
{\bf Abstract}
\end{center}
We argue that Lee-Wick's complex ghost appearing in any higher derivative 
theory is stable and its asymptotic field exists. 
It may be more appropriate to call it ``anti-unstable" in the 
sense that, the more the ghost `decays' into lighter ordinary particles, 
the larger the probability the ghost remains as itself becomes. This is 
explicitly shown by analyzing the two-point functions of the ghost Heisenberg 
field which is obtained as an exact result in the $N\rightarrow\infty$ limit in a massive 
scalar ghost theory with light $O(N)$-vector scalar matter. 
The anti-instability is a 
consequence of the fact that the poles of the complex ghost propagator 
are located on the physical sheet in the complex plane of four-momentum
squared. This should be contrasted to the case of the 
ordinary unstable particle,
whose propagator 
has no pole on the 
physical sheet.

\renewcommand{\thefootnote}{\arabic{footnote}}
\setcounter{footnote}{0}

\newpage

\section{Introduction}

The negative metric ghost mode which is contained in higher derivative 
quantum field theory (QFT) always acquires a complex mass by radiative 
corrections, so becoming   
a complex ghost. Lee and Wick \cite{Lee:1969fy,Lee:1969zze,Lee:1970iw} 
claimed that such a complex ghost cannot be 
created by collisions of positive metric physical particles 
(possessing real energies) because of energy conservation law, so that 
the unitarity of physical particles alone must hold. 

We recently pointed out in our paper \cite{Kubo:2023lpz} 
referred to as I henceforth 
that their treatment of the 
delta function  expressing energy conservation
\begin{equation}
\frac1{2\pi}\int_{-\infty}^\infty dt \, \exp\Bigl(it \sum_i E_i\Bigr)
= \delta_c \Bigl(\sum_iE_i\Bigr) \,,
\end{equation}
which appears at each vertex in a Feynman diagram, 
is {\it wrong} when any of the particles $i$ is a complex ghost possessing 
complex energy $E_i$. This function $\delta_c(E)$ is actually divergent 
when the argument $E$ is complex, but is well-defined as a distribution 
which was first introduced and called complex delta function 
by Nakanishi \cite{Nakanishi:1958,Nakanishi:1958-2} 
 when discussing unstable particles long ago. 
Treating the complex delta function properly, we have shown that 
the complex ghost can actually be created by the collisions 
of physical particles, hence implying the violation of S-matrix unitarity 
of physical particles alone, unfortunately. 

Nevertheless, some people still raise a question: ``Even if the complex 
ghost is created by a collision of physical particles, 
it can decay into lighter 
physical particles so that it eventually disappears after a sufficiently 
long time. Then, the unitarity of physical particles alone is again 
recovered, isn't it?" (See e.g.,
Refs.~\cite{Grinstein:2008bg,Donoghue:2019fcb,Donoghue:2021eto})

To this conceivable question, we have already given a brief but clear answer 
in the paper I. 
We should note that there is a crucial difference between complex ghost 
and ordinary unstable particle; the former has a negative norm while the 
latter a positive norm.  
We have written there as follows: 

``First of all, the ghost state created in the superposition 
$\varphi+\varphi^\dagger$ (i.e., ghost + conjugate-ghost) has 
a negative norm. Since the Dyson's S-matrix of the present system is unitary, 
the negative norm, say $-1$, of the initial ghost state must be conserved. 
So, whatever final states are produced from the initial ghost state, 
the norms of all those final states sum up to the value $-1$ of the initial 
ghost state's norm. To realize this negative value, however, ghost particles 
must be contained among the final states. This implies that the ghosts 
can never disappear by completely `decaying out' into lower mass physical 
particles". 

In contrast, an ordinary unstable particle has a positive norm. Its initial 
norm, say $+1$, 
can be conserved even if it completely decays into ordinary lighter particles and 
disappears, which is indeed the case \cite{Veltman:1963th}.

We think this explanation is enough to prove the {\it stability} of 
the complex ghost particle.  
However, a skeptic might go on to say: 
``A variety of discussions may be possible for the complex ghost 
in the Lee's model. 
However, does such an {\it asymptotic field of complex ghost} really 
exist in the fourth-order derivative theory in the first place?" 

We did not directly answer this objection in the paper I. 
This is because we there discussed the problem solely in the Lee's 
complex ghost model\cite{Lee:1969zze} 
in which the ghost $\varphi$  and conjugate-ghost $\varphi^\dagger$ fields 
are prepared from the beginning. Their asymptotic fields were essentially 
assumed 
to exist in the Lee's model in perturbation theory framework. 
One may cast doubt on the equivalence between Lee's complex ghost 
model and the original fourth-order derivative theory; the ghost in the 
fourth-order 
derivative theory is a single real field while the complex ghost in the 
Lee's model is actually described by two fields, $\varphi$ and $\varphi^\dagger$. 
Is it 
possible at all that such two complex conjugate fields 
emerge from a single real ghost field by radiative corrections?

The purpose of the present paper is to give a clear affirmative answer 
to this problem of existence of the complex (conjugate pair of) ghost 
asymptotic fields. The basic information for asymptotic fields 
contained in a Heisenberg field $\Phi$ is given by the 
two-point Green function (propagator) 
$\VEV{0| \rmT \Phi(x)\Phi(0) |0}$. As this Heisenberg field $\Phi$, we have 
the original fourth-order derivative field $\Phi$ which can be decomposed into 
a positive metric lower-mass field $A$ and the negative metric massive ghost 
field $\phi$; that is, $\Phi=A+\phi$. We have to analyze the system in which 
the massive ghost $\phi$ can get a complex mass by a self-energy diagram
consisting of the loop of light physical particles $\psi_i$. 
However, since the ghost interaction with $\psi_i$ occurs only 
through the original fourth-order derivative field $\Phi=A+\phi$, the same 
self-energy diagram of the $\psi_i$ loop also contribute to $A$-$\phi$ transition 
as well as to $A$'s self-energy. 
For the present problem, however, this mixing between $A$ and $\phi$ fields    
is not essential and 
merely introduces unnecessary complications. So we drop the $A$ component 
field and retain only the ghost field $\phi$ component in $\Phi$ in this 
paper. Also, since we would like to consider a model 
in which our calculation for the ghost two-point function
$\VEV{0| \rmT \phi(x)\phi(0) |0}$  becomes exact in a certain limit, 
we elaborate an $O(N)$-vector scalar matter field model given shortly 
in Section 2.   

We easily compute the ghost two-point function in the leading order 
in $1/N$ expansion. It is merely a one-loop computation but is an exact 
result in the $N \rightarrow\infty$ limit. 
Rewriting the result into the form of the dispersion relation in Section 3, 
and comparing it with the spectral representation,
we can find the asymptotic fields of the system. 
We also derive a sum rule for the spectral function and wave-function 
renormalization factor from the spectral representation in Section 4. 
This sum rule Eq.~(\ref{eq:AntiInst}) may be called anti-instability relation 
and will further solidify the above cited argument 
for the stability of the complex ghost in the paper I. 
Section 5 is devoted to two additional remarks on a confusing point in the 
narrow resonance approximation and on the reason why the ghost asymptotic 
fields appear in a pair of complex conjugate ghosts. In the final Section 6, 
we summarize the results and emphasize the general validity of our argument 
for the existence of asymptotic complex ghost states in any higher 
derivative theories.

\section{The two-point vertex function $\Gamma^{(2)}_\phi(p)$}

We consider the following system of a heavy scalar field $\phi$ with mass $m$ 
and a lighter $O(N)$-vector multiplet of scalar fields $\psi_i$ 
$(i=1,2,\cdots,N)$ 
with mass $\mu\ (<m)$, which is described by the following Lagrangian:
\footnote{Similar model Lagrangians to this Eq.~(\ref{eq:1}) were considered 
in Refs.~\cite{Grinstein:2008bg,Anselmi:2017lia}.}
\begin{equation}
{\cal L}= - \sum_{i=1}^N \half (\partial_\mu\psi_i\partial^\mu\psi_i + \mu^2\psi_i^2)  
-\eps \half (\partial_\mu\phi\partial^\mu\phi+ m_0^2\phi^2)  
+ \sum_{i=1}^N \half \frac{g}{\sqrt N} \phi 
\psi_i\psi_i\,. 
\label{eq:1}
\end{equation} 
Note that we are adopting space-favored metric 
$\eta_{\mu\nu}={\rm diag}(-1, +1,+1,+1)$, so that the $O(N)$-vector light fields 
$\psi_i$ are of positive metric. 
We are interested in the case where the heavier field $\phi$ is a ghost, 
i.e., 
possessing negative metric $\eps=-1$, but, for comparison, we also consider 
the $\eps=+1$ case in which $\phi$ is an ordinary positive metric particle. 

\begin{figure}[htb]
   \centerline{\includegraphics[width=8cm 
               ]{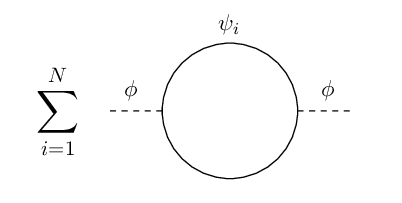}}
 \caption{Ghost self-energy diagram of $\psi_i$-loop in the leading order in 
 $1/N$ expansion.}
 \label{fig:1}
\end{figure}

In the leading order in $1/N$-expansion, the two-point vertex function 
$\Gamma^{(2)}_\phi(p)$ of the heavy field $\phi$ is given by
\begin{equation}
\Gamma^{(2)}_\phi(p)= -\eps(p^2+m_0^2) + \Sigma(p) ,
\end{equation}
where $\Sigma(p)$ stands for the self-energy diagram of $\psi_i$-loop
in Fig.~\ref{fig:1} which reads
\begin{align}
\Sigma(p) &= \half g^2 \int{d^nk\over i(2\pi)^n} \frac1{\mu^2+k^2}\frac1{\mu^2+(p-k)^2} \nn
&= \frac{g^2}{32\pi^2} \int_0^1 dx \left[{\bar\varepsilon\,}^{-1} 
- \ln \bigl(\mu^2+x(1-x)p^2\bigr) \right] \nn
&= \frac{g^2}{32\pi^2} \Bigl[{\bar\varepsilon\,}^{-1} +f(s) \Big]\,, \nn 
f(s) &= 2-\ln \mu^2 - 2\sqrt{\frac{4\mu^2-s}{s}} {\rm Arctan}\sqrt{\frac{s}{4\mu^2-s}} \nn
&= 2-\ln \mu^2 + \sqrt{1-\frac{4\mu^2}{s}} 
\ln \left(\frac{\sqrt{1-{4\mu^2}/{s}} - 1}
{\sqrt{1-{4\mu^2}/{s}} + 1}\right)
\end{align}
with 
\begin{equation}
s\equiv-p^2,\qquad  {\bar\varepsilon}^{-1}\equiv\frac2{4-n} -\gamma+ \ln 4\pi. 
\end{equation}
Here $n$ is the space-time dimension to be eventually set equal to 4.  

The function $f(s)$ in $\Sigma(p)$ has a branch point at $s=4\mu^2$ and we 
take the cut to be along the real axis from $s=4\mu^2$ to $\infty$. The imaginary 
part of $f(s)$ along the cut is given by
\begin{equation}
\lim_{\varepsilon\rightarrow+0} \Im f(s\pm i\varepsilon) = \pm\pi 
\sqrt{1-\frac{4\mu^2}{s}}  \quad \hbox{for real}\ s > 4\mu^2 .
\label{ImaginaryPart}
\end{equation}
Note that the points $s\pm i\varepsilon\ (\varepsilon>0)$ here are taken on the physical sheet 
of the complex $s$ plane. If we take those points on the second sheet, 
the sign $\pm$ on the RHS become opposite sign $\mp$ since 
the upper and lower sides of real $s$ axis for $s>4\mu^2$ on the physical sheet 
smoothly continue to the lower and upper sides on the second sheet, 
respectively.

Depending on the magnitude of the bare mass squared parameter $m_0^2$, 
the two-point function $\Gamma^{(2)}_\phi(p)$ has a zero at $s=m^2$ 
on the real axis of $s (\equiv-p^2)$ below the threshold $s<4\mu^2$, 
or otherwise, has a complex conjugate pair of zeros at 
\begin{equation}
s=
M^2 = m^2 +i\gamma m \quad  \hbox{and}\quad  {M^*}^2 = m^2 -i\gamma m 
\label{eq:6}
\end{equation}
for the $m^2>4\mu^2$ case. That is, $m^2$ is the squared mass of the stable 
particle, or the real part of the complex squared mass of `unstable' particle.
We 
renormalize the mass squared parameter as $m_0^2= m^2 + \delta m^2$ to realize 
\begin{equation}
-\eps \delta m^2 +\Re \Sigma(p)\big|_{s=M^2} = 0 
\end{equation}
(implying $M^2=m^2$ for the case $m^2<4\mu^2$ as shown shortly), 
so that the renormalized two-point vertex function $\Gamma_\phi^{(2)}(p)$ reads 
\begin{align}
\Gamma^{(2)}_\phi(p) = \eps (s-m^2) 
+\frac{g^2}{32\pi^2}\big(f(s)-\Re f(M^2)\big) =:  F(s).
\label{eq:8}
\end{align}
The imaginary part $\Im M^2=\gamma m$ is determined by the requirement that $M^2$ 
be the zero of the two-point vertex function $\Gamma_\phi^{(2)}(p)$: $F(M^2)=0$, i.e.,
\begin{equation}
-\eps (M^2 - m^2) = \frac{g^2}{32\pi^2}\left( f(M^2) - \Re f(M^2) \right)
\ \rightarrow\ 
-\gamma m = \eps \frac{g^2}{32\pi^2} \Im f(m^2+i\gamma m).
\label{eq:9}
\end{equation}
If $m^2<4\mu^2$, since $f(s)$ is real on the real axis of $s$ below the 
threshold $s<4\mu^2$, 
we see that $\gamma=0$ satisfies Eq.~(\ref{eq:9}) and hence 
$-p^2=m^2$ is the zero of the two-point vertex function 
$\Gamma^{(2)}(p)$ so that $m^2$ indeed represents the renormalized 
mass squared of the stable $\phi$-particle, as announced above. However, 
if $m^2$ moves above   
the threshold $4\mu^2$, then $f(s)$ develops the imaginary part 
(\ref{ImaginaryPart}) across the cut and hence 
the zero of $\Gamma^{(2)}(p)$ splits into two zeros 
of a complex conjugate pair, $M^2$ and ${M^*}^2$, as written in Eq.~(\ref{eq:6}). 
As already noted by Lee-Wick \cite{Lee:1969fy,Lee:1969zze,Lee:1970iw}, 
the direction of this splitting is opposite for the ordinary and ghost 
particle cases, $\eps=\pm1$. 
This can be seen explicitly in the present calculation. 
Eq.~(\ref{ImaginaryPart}) implies that the quantity 
$\Im f(m^2+i\gamma m)$ on the RHS of Eq.~(\ref{eq:9}) has the same sign as $\gamma m$ 
when $s=m^2+i\gamma m$ is located on the physical sheet, and has the 
opposite sign to $\gamma m$ when $s=m^2+i\gamma m$ is on the second sheet. 
This means that the solutions $M^2=m^2+i\gamma m$ as well as $M^{*2}$ satisfying 
Eq.~(\ref{eq:9}) exist on the physical sheet only for the ghost case $\eps=-1$, 
while, for the ordinary particle case $\eps=+1$, they move to the second sheet 
and disappear from the physical sheet. 
%

We can give an approximate expression for the complex ghost zero $s=M^2$ 
on the upper-half plane of the physical sheet for the case 
$g/m \lsim O(1)$; then, $(g/m)^2/ 32\pi^2\ll1$ so that 
Eq.~(\ref{eq:9}) implies $\gamma m\ll m^2 \ \rightarrow\  f(m^2 + i\gamma m) 
\simeq f(m^2)$, 
and hence Eq.~(\ref{ImaginaryPart}) leads to
\begin{equation}
M^2 \simeq  
m^2 + i\frac{g^2}{32\pi} \sqrt{1-\frac{4\mu^2}{m^2}} \ .
\end{equation}


\section{Dispersion relation for the $\phi$ propagator}

\begin{figure}[htb]
  \centerline{
   \includegraphics[width=7.5cm]{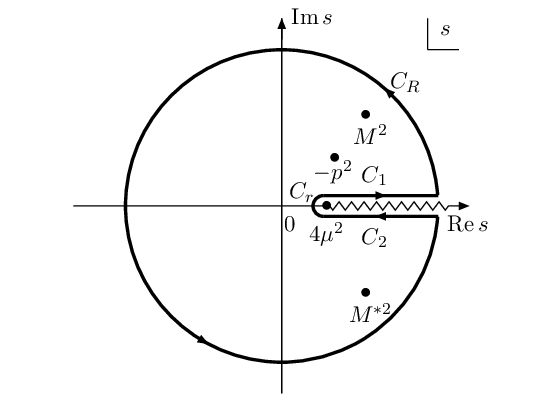}
   }
 \caption{Contour $C=C_1+C_R+C_2+C_r$ on the physical sheet.}
 \label{fig:2}
\end{figure}

Since we have understood the analyticity and singularity structure of the 
two-point function $\Gamma^{(2)}_\phi(p)$, we can now derive a dispersion relation 
for the $\phi$ propagator 
\begin{equation}
D_\phi(s=-p^2)= \frac {i}{\Gamma_\phi^{(2)}(p)} = \frac{i}{F(s)}
\end{equation}
following the usual procedure. 
Consider the following contour integration 
\begin{equation}
I\equiv\frac1{2\pi i}\int_C ds\, \frac{D_\phi(s)/i}{s+p^2} 
\label{Integral}
\end{equation}
(for a general complex value of $-p^2$)
along the closed contour on the physical sheet, 
$ C = C_1+C_R + C_2+C_r$, depicted in Fig.~\ref{fig:2}. 
From the consideration above, 
we know that the propagator $D_\phi(s)$ is the function of $s$ which is real on 
the real axis except on a branch cut starting from $s=4\mu^2$ to $\infty$ and 
analytic everywhere on the physical sheet (i.e., outside the cut) 
other than the complex conjugate poles at $s=M^2$ 
and ${M^*}^2$. So the integrand function $D_\phi(s)/i(s+p^2)$ in 
Eq.~(\ref{Integral}) is also everywhere analytic inside the closed contour 
$C$ except for the three poles at $s=M^2$, ${M^*}^2$ and $s=-p^2$. Applying 
Cauchy's residue theorem, we obtain
\begin{align}
I &=  \frac{D_{\phi}(-p^2)}i  + 
\frac{\eps Z}{M^2+p^2} + \frac{\eps Z^*}{{M^*}^2+p^2}\,, 
\label{eq:I1} \\
Z^{-1} &\equiv\left.\eps \frac{\partial F(s)}{\partial s}\right|_{s=M^2} 
=\lim_{s\rightarrow M^2}\eps \frac{F(s)}{s-M^2}\,.
\end{align}   
On the other hand, since
\begin{align}
&|D_\phi(s)| \rightarrow|s|^{-1} \quad \hbox{as} \quad   |s| \rightarrow\infty, \nn
&|D_\phi(s)| < \exists K \ \ \hbox{on $C_r$ \quad as} \quad r=|s-4\mu^2| \rightarrow0 ,
\end{align}
where $K$ is a finite positive constant, 
the contribution to the integral (\ref{Integral}) 
along $C$ comes only from the discontinuity of the propagator 
across the cut:
\begin{align}
&I = \frac1\pi\int_{4\mu^2}^\infty ds\, \frac{-\rho(s)}{s+p^2} \ ,
\label{eq:I2}\\
& \rho(s) = -\Im \left(\lim_{\varepsilon\rightarrow+0}\frac{D_\phi(s+i\varepsilon)}i\right) 
= -\frac1{2i}\lim_{\varepsilon\rightarrow+0}\left(
\frac1{F(s+i\varepsilon)}-\frac1{F(s-i\varepsilon)}\right) 
=\lim_{\varepsilon\rightarrow+0}\frac{\Im F(s+i\varepsilon)}{|F(s)|^2}\,,
\end{align} 
where $F(s)$ is the two-point function $\Gamma^{(2)}_\phi(p)$ 
as the function of $s$ defined in Eq.~(\ref{eq:8}).
%
Equating Eqs.~(\ref{eq:I1}) and (\ref{eq:I2}), we obtain a dispersion relation 
for our $\phi$ propagator $D_\phi(-p^2)$:\footnote{%
Essentially the same expression as this 
Eq.~(\ref{eq:dispersion}) for the complex ghost propagator was also given by 
Coleman \cite{Coleman:1969xz} and Grinstein et.al. \cite{Grinstein:2008bg}.}
\begin{align}
&{D_{\phi}(-p^2)}  =  
\frac{iZ}{M^2+p^2} + \frac{iZ^*}{{M^*}^2+p^2} 
+\frac1{i\pi}\int_{4\mu^2}^\infty ds  \, \frac{\rho(s)}{s+p^2} 
\label{eq:dispersion}
\\
&\quad 
\hbox{for the complex ghost case; $\Re M^2=m^2>4\mu^2$ and $\eps=-1$} 
\,.
\nonumber
\end{align}

This Eq.~(\ref{eq:dispersion}) takes the form of K\"allen-Lehman's spectral representation for the 
propagator, so, if we recall the well-known method of its derivation 
by inserting 
the complete set of states as intermediate states in 
the operator expression of the 
two point function\footnote{%
Here in Eq.~(\ref{Hpropagator}), the $p^0$-integration must be performed 
along a much deformed contour $C(p^0)$ from the real axis on the complex $p^0$ 
plane while 3d $\bfp$-integration 
is the usual Fourier transformation along the real axis of $\bfp$, as will 
be explained at the end of this section.  
}%
\begin{equation}
 \VEV{0| {\rm T} \phi(x) \phi(0) |0}
\left(=\int_{C(p^0)}\frac{d^4p}{(2\pi)^4} e^{ipx} D_\phi(-p^2)\right)\,,
\label{Hpropagator}
\end{equation}
we can understand the meaning of each term on the RHS of 
Eq.~(\ref{eq:dispersion}). 
The last $\rho(s)$ integral term is understood, as usual, as coming from the 
discontinuity caused by 
the continuum spectrum of two physical $\psi_i$-particle intermediate state. 
Then, the first and second pole terms must be understood as coming from the 
two discrete one-particle states possessing complex conjugate masses $M^2$ and 
${M^*}^2$. 
Note that {\it the poles appearing on the physical sheet mean 
the existence of the corresponding one-particle asymptotic states in the 
complete set of states of the theory.} 

Indeed, it is instructive to consider 
the same propagator $D_\phi(-p^2)$ for the other parameter-value cases 
in the present system (\ref{eq:1}).  
First consider the case $m^2<4\mu^2$, for which $D_\phi(s)$ has only a single 
pole term at $s=m^2$ on the real axis, so that the dispersion relation 
(\ref{eq:dispersion}) takes the form
\begin{equation}
{D_{\phi}(-p^2)}  =  
\frac{\eps}i \frac{Z}{m^2+p^2} 
+\frac1{i\pi}\int_{4\mu^2}^\infty ds\, \frac{\rho(s)}{s+p^2}
\qquad \hbox{for $m^2<4\mu^2$ case}\,. 
\label{eq:dispersion2}
\end{equation}
This is the usual {\it stable particle} 
case if $\eps=+1$: There is a one-particle pole in the propagator $D_\phi(s)$ and 
the corresponding asymptotic field $\phi^{\rm as}(x)$ exists 
satisfying $(\square-m^2)\phi^{\rm as}(x)=0$.
Next consider a more interesting case, $m^2>4\mu^2$ with $\eps=+1$ 
(i.e., positive norm). This is the ordinary {\it unstable particle} case, 
for which the complex conjugate poles move into the second sheet and 
disappear from the physical sheet, as explicitly shown above, so that 
the dispersion relation (\ref{eq:dispersion}) takes the form
\begin{equation}
{D_{\phi}(-p^2)}  =  
+\frac1{i\pi}\int_{4\mu^2}^\infty ds\, \frac{\rho(s)}{s+p^2}
\qquad \hbox{for $m^2>4\mu^2$ case with $\eps=+1$}\,. 
\label{eq:dispersion3}
\end{equation}
There is no one-particle pole term here, which agrees with the fact that 
there exists no asymptotic field corresponding to an unstable particle.
As everyone knows, however small the decay probability (into two $\psi_i$ 
particle states here) is, any unstable particle decays out 
into lighter ordinary particles and eventually disappears in 
sufficiently long time \cite{Veltman:1963th}.  
So the complete set of states is spanned by stable particles alone.

      
We thus conclude from the dispersion relation (\ref{eq:dispersion}) 
for the $\phi$ propagator 
that the Heisenberg field $\phi$ (massive regulator part 
of the fourth-order derivative Heisenberg field) has the conjugate 
pair of asymptotic fields of complex ghost, $\varphi$ and $\varphi^\dagger$:
\begin{equation}
\phi(x) \ \  \myLlimit{x^0 \rightarrow\infty} Z^{1/2} \varphi(x) + {Z^*}^{1/2} \varphi^\dagger(x),\label{AsymptoticF} 
\end{equation}
which satisfy the free field equations, 
$(\square-M^2)\varphi(x)=0$ and its complex conjugate 
$(\square-M^{*2})\varphi^\dagger(x)=0$.

\begin{figure}[htb]
  \centerline{
   \includegraphics[width=7.5cm]{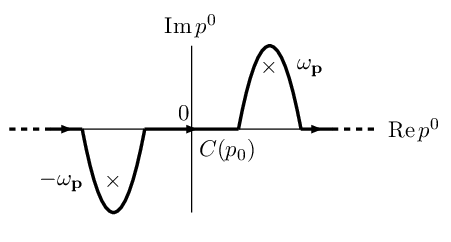}
   }
 \caption{$p^0$ integration contour $C(p^0)$ in Eq.~(\ref{GhostProp}).}
 \label{fig:3}
\end{figure}

The unfamiliar metric structure of these 
asymptotic complex fields can most easily be found 
simply by canonical quantization of their unique free field 
Lagrangian:\footnote{%
The overall sign of this action is a convention which can be changed by 
redefining the asymptotic field $\varphi$ to $i\varphi$. We use 
the same sign choice as in our previous paper I, which is opposite 
to Nakanishi's in Ref.~\cite{Nakanishi:1972wx}.}   
\begin{equation}
\calL = 
\frac12\left[\partial_\mu\varphi\,\partial^\mu\varphi + M^2\varphi^2  
+\partial_\mu\varphi^\dagger\,\partial^\mu\varphi^\dagger+ {M^*}^2{\varphi^\dagger}^2  \right]. 
\label{eq:GhostL}
\end{equation}
But we note that this is essentially the same Lagrangian as given by 
Nakanishi \cite{Nakanishi:1972wx} for the 
BC field sector of the Lee's complex ghost model \cite{Lee:1969zze} 
in which 
the complex ghost fields 
$\varphi$ and $\varphi^\dagger$, or equivalently, 
$B=(\varphi +\varphi^\dagger)/\sqrt2$ and    
$C=i(\varphi -\varphi^\dagger)/\sqrt2$ fields, are not the 
asymptotic fields but the fields 
introduced in the model from the beginning. 
Anyway, since the property of these fields is uniquely specified by 
the Lagrangian (\ref{eq:GhostL}), we can use the Nakanishi's results which 
we recapitulated in our previous paper I. The complex ghost field $\varphi(x)$ 
is expanded into plane waves as 
\begin{equation}
\varphi(x) = \int\frac{d^3\bfp}{\sqrt{(2\pi)^32\omega_{\bfp}}}
\left(
\alpha(\bfp)e^{i\bfp\bfx-i\omega_{\bfp}x^0} 
+ \beta^\dagger(\bfp)e^{-i\bfp\bfx+i\omega_{\bfp}x^0}
\right)\,,
\label{eq:varphi}
\end{equation}
where $\omega_{\bfp}$ is the complex energy 
$\omega_{\bfp} = \sqrt{ \bfp^2 + M^2 }$
and the creation and annihilation operators satisfy the off-diagonal 
commutation relations:
\begin{eqnarray}
&&[\alpha(\bfp), \beta^\dagger(\bfq)] = [\beta(\bfp), \alpha^\dagger(\bfq)]= -\delta^3(\bfp- \bfq)\,, \nn
&&[\alpha(\bfp), \alpha^\dagger(\bfq)] =[\beta(\bfp), \beta^\dagger(\bfq)]= 0\,.
\label{eq:CCR}
\end{eqnarray}
Then, by using these, the Feynman propagator is immediately found to be 
given as
\begin{align}
&\bra0 \rmT \varphi(x)\,\varphi(y) \ket0 \nn
&=
-\int{d^3\bfp\over(2\pi)^32\omega_\bfp} 
\Bigl\{
\theta(x^0-y^0) e^{i\bfp(\bfx-\bfy)-i\omega_\bfp (x^0-y^0)} 
+\theta(y^0-x^0) e^{-i\bfp(\bfx-\bfy)+i\omega_\bfp(x^0-y^0)} 
\Bigr\}\,.
\label{varphiProp}
\end{align} 
This 3d momentum expression is rewritten into 4d momentum expression by 
introducing $p^0$ variable as  
\begin{equation}
= -\int_{C(p^0)} \frac{d^3\bfp\,dp^0}{i(2\pi)^4} \,
\frac{e^{i\bfp(\bfx-\bfy)-ip^0(x^0-y^0)}}{M^2+p^2}\,.
\label{GhostProp}
\end{equation} 
In order for this 4d expression to reproduce the 3d expression 
(\ref{varphiProp}), 
the $p^0$ integration contour 
$C(p^0)$ here has to be the deformed one from the real axis $R=(-\infty,+\infty)$ 
such that it passes below the left pole at 
$p^0=-\omega_\bfp$ and above the right pole at $p^0=+\omega_\bfp$ as shown in 
Fig.~\ref{fig:3}. 
With this understanding, we see that the asymptotic fields 
$Z^{1/2}\varphi + Z^{*\,1/2}\varphi^\dagger$ actually reproduce the 
two complex conjugate poles 
in the $\phi$-propagator (\ref{eq:dispersion}) in 4d momentum representation. 

\section{Spectral representation for the commutator}

Once a spectral representation is found for one type of two-point function, 
we can immediately write down those for other type of two-point functions 
by using the same spectral function. 
We first note that the dispersion relation (\ref{eq:dispersion}) is rewritten 
into the spectral representation for the propagator in $x$ space:
\begin{equation}
\VEV{0|\,\rmT \phi(x)\phi(0)\,|0} 
= -Z \Delta_\rmF(x; M^2) - Z^*\Delta_\rmF(x; {M^*}^2)
+ \int_{4\mu^2}^\infty ds\,\frac{\rho(s)}\pi\Delta_\rmF(x; s)\,, 
\end{equation} 
where $\Delta_\rmF(x; m^2)$ denotes the Feynman propagator function 
for the 
free field with mass squared $m^2$ including also the complex $m^2$ case:
\begin{equation}
\Delta_\rmF(x; m^2) = \int_{C(p^0)}{d^4p\over i(2\pi)^4}\,\frac{e^{ipx}}{m^2+p^2}\,.
\end{equation}

Knowing this form, we can immediately write down, 
in particular, the VEV of the commutation relation of the 
Heisenberg operator $\phi$, in which we are now interested:
\begin{equation}
\VEV{0|\, [\phi(x),\, \phi(0)] \,|0}
= -Z i\Delta(x; M^2) - Z^* i\Delta(x; {M^*}^2)
+ \int_{4\mu^2}^\infty ds\,\frac{\rho(s)}\pi i\Delta(x;s)\,, 
\label{eq:CRvev}
\end{equation}
in terms of the commutator function $i\Delta(x; m^2)$ 
for the free field with (generally complex) mass squared 
$m^2$: \cite{Nakanishi:1972wx}
\begin{equation}
\Delta(x; m^2) = \int{d^3\bfp\over(2\pi)^3E_\bfp} \sin[ \bfp\cdot\bfx -E_\bfp x^0 ], 
\qquad E_\bfp=\sqrt{\bfp^2+ m^2}\,.
\end{equation}

This Eq.~(\ref{eq:CRvev}) leads to a very interesting relation. Take a time derivative 
$\partial/\partial x^0$ and set $x^0=0$ on both sides of this equation. Then, 
the LHS is reduced to the equal time commutator between  the 
Heisenberg field 
operator $\phi(0)$ and its conjugate momentum operator 
$\pi(x) \equiv\partial{\calL}/\partial\dot\phi(x)$ at time $x^0=0$, which yields a simple 
value thanks to the canonical commutation relation (CCR):
\begin{equation}
\hbox{LHS}=\bra0 [\dot\phi(\bfx, 0),\, \phi(0)] \ket0 = 
\VEV{0|\,[\eps \pi(\bfx,0),\, \phi(0)]\,|0} = -i \eps \delta^3(\bfx)\,.
\end{equation}
On the RHS also, since $i\Delta(x; m^2)$ is the free field commutator function, 
the CCR gives the same quantity 
\begin{equation}
i\dot\Delta(\bfx, 0; m^2) = -i \delta^3(\bfx)
\end{equation}
independently of the mass value $m^2$. Using this and dividing both sides 
by a common factor $-i \delta^3(\bfx)$, we obtain
\begin{equation}
-1 = -(Z + Z^*) + \int_{4\mu^2}^\infty ds\ \frac{\rho(s)}\pi 
\qquad \hbox{for ghost field case}
\, .
\label{eq:ghost}
\end{equation}
Note that we have set $\eps = -1$ on the LHS here since this relation 
is derived from the dispersion relation 
(\ref{eq:dispersion}) valid for the ghost field case. 
If we apply the same procedure to the dispersion relations, 
Eqs.~(\ref{eq:dispersion2}) 
and (\ref{eq:dispersion3}), for ordinary stable and unstable particle cases, 
respectively, we obtain 
\begin{align}
+1 &= Z + \int_{4\mu^2}^\infty ds\ \frac{\rho(s)}\pi 
\qquad \hbox{for stable particle case with $\eps=+1$}\,, 
\label{eq:stable}\\ 
+1 &=  \int_{4\mu^2}^\infty ds\ \frac{\rho(s)}\pi 
\qquad \hbox{for unstable particle case}\,.
\label{eq:unstable}
\end{align}
Eq.~(\ref{eq:stable}) is the well-known relation written in any field theory 
textbook, whose physical interpretation following from the derivation of 
spectral representation is as follows: 
$Z$ is the probability that the state $\phi(x)\ket{0}$ (generated by acting 
the Heisenberg field $\phi(x)$ on the vacuum $\ket{0}$) contains the one-particle 
state $\ket{\bfp; m^2}$, while 
$$
\int_{4\mu^2}^\infty ds\ \frac{\rho(s)}{\pi} =: c>0
$$ 
represents the probability that the state $\phi(x)\ket{0}$ contains 
the many particle states (only two particle states in this calculation). 
So the Eq.~(\ref{eq:stable}) says that the total probability 
that $\phi(x)\ket{0}$ contains one-particle and many particle states 
adds up to 1. 
In the same way, Eq.~(\ref{eq:unstable}) for the unstable particle case 
shows that $\phi(x)\ket{0}$ contains 
no one-particle asymptotic state and the total probability is saturated 
only by the contribution $c$ from the continuum many particle states 
consisting of 
lighter particles produced by decays.

Now comes the relation (\ref{eq:ghost}), in view of which we reach the 
following interpretation.
First, $Z+Z^*$ represents the probability 
that $\phi(x)\ket{0}$ contains the complex ghost asymptotic one-particle state, 
which is 
the {\it superposition of ghost $\varphi(x)\ket{0}$ and conjugate ghost} 
$\varphi^\dagger(x)\ket{0}$ as explained above. The state $\phi(x)\ket{0}$ also 
contains many particle states which appear as the `decay products' 
of the 
original ghost $\phi$. The probability of the many particle states 
of `decay products', 
$\int_{4\mu^2}^\infty ds\ \rho(s)/\pi= c >0$, is the same as the previous two cases 
and hence positive.  Then, the relation (\ref{eq:ghost}) tells us a very 
interesting but counter-intuitive relation:\footnote{Exactly the same equation 
as this Eq.~(\ref{eq:AntiInst}) was written in Eq.~(72) in Coleman's lecture 
\cite{Coleman:1969xz}, where he noted that it indicates the high-$p^2$ behavior 
of the original propagator in the fourth-order derivative theory remains as 
good as that at tree level, i.e., damps as $\sim1/p^4$.}
\begin{equation}
Z+Z^* = 1+c\,.
\label{eq:AntiInst}
\end{equation}
Surprisingly, the probability $Z+Z^*$ 
that the ghost remains as itself 
becomes even larger as  $c$ increases;  that is, the more the ghost `decays' 
into lighter ordinary particles, the larger the probability the ghost remains 
as itself becomes. 
But this strange property of ghost, which we call 
{\it anti-instability},\footnote{This word is inspired by 
Coleman's lecture in Ref.~\cite{Coleman:1969xz} in which he suggested that 
the complex ghosts be referred to as ``antistable particles" because of 
{\it the radical difference} of situations 
from the ordinary unstable particles.}  
was already pointed out in our previous paper I, as cited in the introduction. 

We thus have completely proved the existence of asymptotic 
fields of complex ghost, $\varphi$ and $\varphi^\dagger$ 
in Eq.~(\ref{AsymptoticF}). It is guaranteed by the anti-instability of the 
negative metric ghost.

\section{Two additional remarks}

Before closing this paper, we add two remarks.

One is on a possibly confusing point concerning the narrow resonance 
approximation given by Grinstein et.al. in Ref.~\cite{Grinstein:2008bg}.
Those authors also wrote down the same form of dispersion relation as 
our Eq.~(\ref{eq:dispersion}) for the complex ghost field in a similar 
scalar field theory model. They noticed that the spectral density $\rho(s)$ is 
approximately given by
\begin{equation}
\rho(s)\simeq  \frac{m\gamma}{(s-m^2)^2+m^2\gamma^2}
=\frac1{2i}\left[
\frac1{s-m^2-im\gamma}-\frac1{s-m^2+im\gamma}
\right] 
\end{equation}
near the resonant energy $s=m^2$ for narrow resonance case $\gamma\ll m$. 
This is generally true 
since the condition $\gamma/m \ll1$ (implying $(g/m)^2/32\pi\ll1$ in our case)  
anyway means the small coupling constant so that $Z\simeq 1$ and 
$1/F(s+i\varepsilon)\simeq -Z/(s-M^2)\simeq -1/(s-M^2)$ and 
$1/F(s-i\varepsilon) 
\simeq -1/(s-M^{*2})$ near the energy $s=m^2$.  
Noting that $\rho(s)$ is strongly peaked at $s=m^2$, 
they extend the 
$s$-integration in Eq.~(\ref{eq:dispersion}) over the whole real axis $R$ 
of $s$. 
Then the $s$-integration can be done by closing the integration contour 
in either upper or lower half-plane and yields 
\begin{align}
\frac1{i\pi}\int_{-\infty}^\infty ds  \, \frac{\rho(s)}{s+p^2 }
&= - \frac{i}{M^{*2}+p^2}  \qquad \hbox{for}\ \ \Im(-p^2)>0
\end{align}
for the $-p^2=(p^0)^2-\bfp^2$ variable on the {\it upper half plane}.
Then, if this is substituted into the third 
continuum term in 
Eq.~(\ref{eq:dispersion}), it cancels(!) the second pole term 
$iZ/(M^{*2}+p^2)$ in this approximation with $Z\simeq 1$, 
and Eq.~(\ref{eq:dispersion}) now gives the expression for the 
ghost propagator 
\begin{equation}
{D_{\phi}(-p^2)}  \simeq   
\frac{i}{M^2+p^2}  \qquad \hbox{for}\ \ \Im(-p^2)>0.
\label{eq:upper}
\end{equation}
In the same way, for $-p^2$ on the lower half plane, we find 
\begin{equation}
{D_{\phi}(-p^2)}  \simeq   
\frac{i}{M^{*2}+p^2} \qquad \hbox{for}\ \ \Im(-p^2)<0.
\label{eq:lower}
\end{equation}
These are of course valid approximate results 
for $-p^2$ on the upper and lower half planes, respectively, but 
nevertheless rather misleading expressions. 
It should never be interpreted that 
they imply the disappearance of the complex ghost pole at $-p^2=M^2$ 
or the complex conjugate ghost pole at $-p^2=M^{*2}$. 
They are merely the approximate results numerically valid only 
near $-p^2=m^2$ 
above the cut $\Re(-p^2)\geq4\mu^2$. 
For instance, for real $-p^2$ below the threshold $-p^2<4\mu^2$, 
$D_\phi(-p^2)$ is real (and has no cut) since both complex conjugate poles 
contribute to it, but the approximate result, either Eq.~(\ref{eq:upper})
or Eq.~(\ref{eq:lower}), is complex, so failing in reflecting the analytic 
structure of the propagator function $D_\phi(-p^2)$. 
Eqs.~(\ref{eq:upper}) and (\ref{eq:lower}) are two separate functions 
and either expression, (\ref{eq:upper}) or 
(\ref{eq:lower}), does not know the existence of the pole at the other 
half-plane. 

Another remark is on the reason why the asymptotic fields appear in a pair of 
complex conjugate ghosts, $\varphi$ and $\varphi^\dagger$. 
One obvious reason is that the original Heisenberg field $\phi$ 
is a real field. The asymptotic field of the hermitian Heisenberg field should 
be real as a whole, so the combination 
\begin{equation}
Z^{1/2}\varphi(x)+{Z^*}^{1/2}\varphi^\dagger(x)\equiv|Z|^{1/2}\phi^{\rm as}(x). 
\label{eq:superpose}
\end{equation}
Another reason is to realize the non-vanishing and negative norm of the 
original Heisenberg field $\phi$. As is proved generally and easily, 
any complex energy eigenstate of hermitian Hamiltonian is of zero-norm and 
can have non-vanishing innerproduct only with its conjugate energy eigenstate.
The superposition state created by the combination (\ref{eq:superpose}) of 
conjugate pair of asymptotic fields are just such a state that can carry 
non-vanishing (negative) norm; indeed, using the {\it real} asymptotic 
field $\phi^{\rm as}$ in Eq.~(\ref{eq:superpose}) and 
inserting the plane wave expansion 
(\ref{eq:varphi}) and its complex conjugate for the asymptotic fields 
$\varphi$ and $\varphi^\dagger$ there, we have 
\begin{align}
\ket{ \bfp; x^0} &\equiv 
\int d^3\bfx\,\sqrt{\frac{2|\omega_\bfp|}{(2\pi)^3}}\,
e^{i\bfp \bfx}\,\phi^{\rm as}(x)\ket0  \nn
&= \left(e^{i(\theta_Z-\theta_\bfp)/2}\beta^\dagger(\bfp) e^{i\omega_\bfp x^0}
+e^{-i(\theta_Z-\theta_\bfp)/2}\alpha^\dagger(\bfp) e^{i\omega^*_\bfp x^0}\right)\ket0\,,
\label{eq:realstate} 
\end{align}
where $\theta_Z$ and $\theta_\bfp$ are the phases of $Z$ and $\omega_\bfp$; 
$Z = |Z|e^{i\theta_Z}$ and $\omega_\bfp= |\omega_\bfp| e^{i\theta_\bfp}$.  
The norm of this superposition is calculated by using the commutation relations 
(\ref{eq:CCR}) 
as
\begin{align}
\VEV{\bfq; x^0 | \bfp; x^0} 
&= \bra0 [\alpha(\bfq),\,\beta^\dagger(\bfp)] e^{i(2\theta_Z-\theta_\bfp-\theta_\bfq)/2} 
e^{i(\omega_\bfp-\omega_\bfq)x^0}\ket0 \nn
& \hspace{1em}{}+ \bra0 [\beta(\bfq),\,\alpha^\dagger(\bfp)] 
e^{-i(2\theta_Z-\theta_\bfp-\theta_\bfq)/2} e^{i(\omega^*_\bfp-\omega^*_\bfq)x^0}\ket0 \nn
&= -\delta^3(\bfq-\bfp)\, 2\cos (\theta_Z -\theta_\bfp) \,.
\label{eq:Norm}
\end{align}
This realizes the non-vanishing negative norm as announced above.\footnote{%
It can easily be proved at least for the case $\mu^2/m^2\ll1$ that the phase 
condition $0<\theta_Z -\theta_\bfp<\pi/2$ holds independently of 
the coupling strength $g^2/m^2$, so that the norm (\ref{eq:Norm}) 
always remains negative.}
 One should 
also note the fact that {\it this norm of the state $|\bfp; x^0\rangle$ 
generated by the real asymptotic field $\phi^{\rm as}(x)$ is independent of 
time $x^0$}.
This is remarkable since the first and second states in Eq.~(\ref{eq:realstate}) which 
are created by the complex asymptotic fields $\varphi$ and $\varphi^\dagger$, 
respectively, each has a terrible time dependence, 
$e^{i\omega_\bfp x^0}$ or $e^{i\omega^*_\bfp x^0}$, exponentially 
divergent or damping as $x^0\rightarrow\pm\infty$. These time-dependent coefficients are, 
actually, fake since they are irrelevant to the magnitude of the state 
$\beta^\dagger(\bfp)\ket0$ nor $\alpha^\dagger(\bfp)\ket0$ because they are of zero norm.    
Those exponentially divergent and damping factors cancel each other between 
$\beta^\dagger(=\varphi)$ and $\alpha^\dagger(=\varphi^\dagger)$ states and realize an 
$x^0$-independent state. In this sense, it is very important that 
the asymptotic fields of $\phi$ always appear in the real combination 
(\ref{eq:superpose}) which creates the superposition states 
(\ref{eq:realstate}).

\section{Conclusion}

We have shown in this paper that the 
complex conjugate pair of ghost fields $\varphi$ and $\varphi^\dagger$ 
are actually contained as asymptotic fields in the ghost (regulator) 
Heisenberg field $\phi$ in the higher derivative theories. 
Those asymptotic complex conjugate ghost particles each have zero norm 
but always appear in a superposition form of ghost and conjugate ghost, 
and carries negative norm. Owing to the negative norm, they have a peculiar 
stability property called anti-instability; the more they `decay' into 
ordinary particles, the more the ghosts appear. 
This solidifies our previous conclusion that the unitarity of 
physical particles alone is necessarily violated as far as 
(negative metric) ghost exists in the theory.

This anti-instability is concisely expressed in the form of 
Eq.~(\ref{eq:AntiInst}); $Z+Z^* = 1 + c$, where 
$Z+Z^*$ is the probability of 
complex ghost one-particle state and $c$ the probability of many ordinary 
particle states contained in the state $\phi(x)\ket0$ created by the 
ghost Heisenberg field $\phi(x)$.  
It was proved based on the dispersion relation 
Eq.~(\ref{eq:dispersion}) for the ghost propagator. 
The proof is very robust and non-perturbative in the sense that 
the anti-instability 
relation of the form (\ref{eq:AntiInst}) can always be 
derived as far as the analyticity property of the ghost propagator 
is given as shown in Fig.~\ref{fig:2}. We only needed the positivity of the 
discontinuity function $\rho(s)$ along the cut on the real axis. 

We should, however, also note the fact that the analyticity structure 
as shown in Fig.~\ref{fig:2} is very special. Once the intermediate states 
contain the complex ghost particle, their energy $P^0$ takes values 
extending over two-dimensional region on the complex $P^0$ plane, i.e., 
not restricted on the (one-dimensional) real axis, and so the 
dispersion relation or the spectral representation would become 
much more complicated form whose precise expression has never been 
given (cf.~Ref.~\cite{Anselmi:2017lia}).

Nevertheless, on the other hand, this also implies the robustness of our 
general conclusion that the physical $S$-matrix unitarity is violated 
in higher 
derivative theories. The fields in those theories are always decomposed into 
second-order derivative fields among which some massive fields are 
negative metric ghosts. Our discussions in this paper 
can apply to those ghost fields, which necessarily become complex ghosts 
by the `decay' to the ordinary lighter particles.  
Assume that the unitarity with ordinary positive metric particles 
alone could hold. 
Then, the total state vector space must be spanned only 
with those ordinary particles. If so, we have the usual form of spectral 
representation for the two-point functions of the ghost Heisenberg field 
$\phi$, which are given by the integral along real $s$ axis with 
positive definite spectral function $\rho(s)$. In particular, from the  
spectral representation for the function $\bra0 [\phi(x),\, \phi(0)] \ket0$, 
we would get the anti-instability relation (\ref{eq:AntiInst}) 
with no complex ghost asymptotic states, $Z+Z^*=0$, so that $1+c=0$. 
But this contradicts the positivity assumption of $\rho(s)$, 
$c=\int ds \rho(s) >0 $. So the original unitarity assumption is wrong. 

This implies that there must exist the complex ghost states in the total 
state vector space and those ghost states are contained in the state 
$\phi(x)\ket0$ of the ghost Heisenberg field $\phi(x)$. 
So we must admit the fact that the complex ghosts are necessarily created via 
the ghost field $\phi(x)$ contained in the original higher derivative 
field. If we could have the physical unitarity despite this, therefore, 
the only possibility would be to find (or construct) a special 
higher derivative theory possessing a certain symmetry like BRST symmetry 
in gauge theories 
which guarantees that the complex ghosts always appear in zero-norm 
combinations in the physical subspace specified by the charge of the symmetry.


\section*{Acknowledgment}

We thank Bob Holdom for critical discussions and, in particular, for raising 
questions whether the asymptotic complex ghost states exist or not, which 
motivated the research of this paper.
We also thank Noboru Nakanishi for his critical but helpful comments on 
our work. 
This work was supported in part by the MEXT/JSPS KAKENHI Grant Number
JP18K03659 (T.K.) and 23K03383 (J.K.).

\let\doi\relax


\begin{thebibliography}{10}

\bibitem{Lee:1969fy}
T.~D. Lee and G.~C. Wick, Nucl. Phys. B, {\bf 9}(2), 209 (1969).

\bibitem{Lee:1969zze}
T.~D. Lee,
\newblock {\it A relativistic complex pole model with indefinite metric}, in
  {\it Quanta: Essays in Theoretical Physics Dedicated to Gregor Wentzel} (Chicago
  University Press, Chicago, 1970), p. 260.

\bibitem{Lee:1970iw}
T.~D. Lee and G.~C. Wick, Phys. Rev. D, {\bf 2}(6), 1033 (1970).

\bibitem{Kubo:2023lpz}
J.~Kubo and T.~Kugo,
PTEP \textbf{2023} (2023) no.12, 123B02
doi:10.1093/ptep/ptad143
[arXiv:2308.09006 [hep-th]].

\bibitem{Nakanishi:1958}
N.~Nakanishi, {Prog. Theor. Phys.}, {\bf 19}, 607 (1958).

\bibitem{Nakanishi:1958-2}
N.~Nakanishi, {Prog. Theor. Phys.}, {\bf 20}, 822 (1958).

\bibitem{Grinstein:2008bg}
B.~Grinstein, D.~O'Connell, and M.~B. Wise, Phys. Rev. D, {\bf 79},
  105019 (2009),  [arXiv:0805.2156[hep-th]].

\bibitem{Donoghue:2019fcb}
John~F. Donoghue and Gabriel Menezes, Phys. Rev. D, {\bf 100}(10), 105006
  (2019),  [arXiv:1908.02416[hep-th]].

\bibitem{Donoghue:2021eto}
John~F. Donoghue and Gabriel Menezes, Phys. Rev. D, {\bf 104}(4), 045010
  (2021),  [arXiv:2105.00898[hep-th]].

\bibitem{Veltman:1963th}
M.~J.~G.~Veltman,
Physica \textbf{29} (1963), 186-207
doi:10.1016/S0031-8914(63)80277-3

\bibitem{Anselmi:2017lia}
D.~Anselmi and M.~Piva,
Phys. Rev. D \textbf{96} (2017) no.4, 045009
doi:10.1103/PhysRevD.96.045009
[arXiv:1703.05563 [hep-th]].
%

%
%
%
%
%
%
%
\bibitem{Coleman:1969xz}
S.~Coleman,
\newblock {\it Acausality},
\newblock in {\it {7th International School of Subnuclear Physics (Ettore
  Majorana): Subnuclear Phenomena}} (1969), p282.
%
\bibitem{Nakanishi:1972wx}
N.~Nakanishi, Phys. Rev. D, {\bf 5}, 1968 (1972).
%
%
%
%
%
%
%
%
%
%
%
%
%
%

\end{thebibliography}

\end{document}